\documentclass[fleqn,10pt]{wlscirep}
\usepackage[utf8]{inputenc}

\usepackage{times}

\usepackage{csquotes}
\usepackage{soul}
\usepackage{url}
\usepackage{amsmath}
\usepackage{amsthm}
\usepackage{booktabs}
\usepackage{algorithm}
\usepackage{algorithmic}
\usepackage{booktabs}
\usepackage{multirow}
\usepackage{multirow}
\usepackage[T1]{fontenc}

\usepackage[T1]{fontenc}
\usepackage{setspace}
\theoremstyle{definition}

\def\ie{{\em i.e.},\ }
\def\eg{{\em e.g.},\ }

\title{Prediction of Tissue Outcome in Acute Ischemic Stroke using imbalanced Temporal Deep Gaussian Process (iTDGP)}

\author[1,*]{Mohsen Soltanpour}
\author[1]{Muhammad Yousefnezhad}
\author[1,3]{Russ Greiner}
\author[1]{Pierre Boulanger}
\author[2]{Brian Buck}

\affil[1]{University of Alberta, Department of Computing Science, Alberta, Canada}
\affil[2]{University of Alberta, Department of Medicine, Alberta, Canada}

\affil[2]{Alberta Machine Intelligence Institute, Alberta, Canada}

\affil[*]{msoltanp@ualberta.ca}

\affil[*]{Corresponding Author}


\begin{abstract}
As one of the leading causes of mortality and disability worldwide, Acute Ischemic Stroke (AIS) occurs when the blood supply to the brain is suddenly interrupted because of a blocked artery. Within seconds of AIS onset, the brain cells surrounding the blocked artery die, which leads to the progression of the lesion. The automated and precise prediction of the existing lesion plays a vital role in the AIS treatment planning and prevention of further injuries. The current standard AIS assessment method, which thresholds the 3D measurement maps extracted from Computed Tomography Perfusion (CTP) images, is not accurate enough. Due to this fact, in this article, we propose the imbalanced Temporal Deep Gaussian Process (iTDGP), a probabilistic model that can improve AIS lesions prediction by using baseline CTP time series. Our proposed model can effectively extract temporal information from the CTP time series and map it to the class labels of the brain's voxels. In addition, by using batch training and voxel-level analysis iTDGP can learn from a few patients and it is robust against imbalanced classes. Moreover, our model incorporates a post-processor capable of improving prediction accuracy using spatial information. Our comprehensive experiments, on the ISLES 2018 and the University of Alberta Hospital (UAH) datasets, show that iTDGP performs better than state-of-the-art AIS lesion predictors, obtaining the (cross-validation) Dice score of $71.42\%$ and $65.37\%$ with a significant $p<0.05$, respectively.

\end{abstract}
\begin{document}

\flushbottom
\maketitle
%
%
\thispagestyle{empty}

\section{Introduction} 
Acute Ischemic Stroke (AIS), accounting for almost 90\% of all strokes, is a sudden interruption of blood flow in a blood vessel supplying the brain~\cite{smith2018accuracy}. Each year, AIS claims approximately 6.2 million lives, which is more than the number of deaths caused by AIDS, tuberculosis, and malaria combined~\cite{alwan2011global}. Treatment of AIS is a highly time-sensitive task since irreversible brain damage can occur if blood flow is not restored promptly~\cite{goyal2016endovascular}. A major step leading to the best AIS treatment decision is predicting future AIS lesions (\ie the brain regions that will be damaged if blood flow is not restored quickly) by using the baseline imaging data~\cite{christensen2019ct}. The most common baseline imaging method for AIS assessment is Computed Tomography Perfusion (CTP), in which a contrast bolus is injected into the patient's bloodstream and a CT scan of the patient's brain is taken every second as the bolus passes through the patient's arteries~\cite{konstas2009theoretic}. 

The current standard approach to predicting the AIS lesion is simply thresholding the CTP parameter maps, which are predefined features extracted from CTP time series to quantify brain cerebral circulation~\cite{laughlin2019rapid}. Unfortunately, this simple method does not accurately predict the lesion volumes, which can lead to poor treatment decisions~\cite{nielsen2018prediction}. AIS lesion prediction task is challenging for four main reasons: (1)~we need to predict the future lesion --- \ie this study investigates both short-term (within 3 hours of baseline CTP imaging) and long-term (after $24(\pm 6)$ hours of baseline CTP imaging) predictions for future AIS lesions --- based only the baseline information; (2)~the small difference between the appearance of lesion and the rest of the brain tissues in the CTP images; (3)~the heterogeneity of the lesion in size, shape, and loci (across different stroke patients); and (4)~the small number of labelled (actual future lesions) brain image samples. 
Recent developments in deep neural networks have led to many attempts at improving the accuracy of AIS lesion prediction. These approaches can be grouped into two main categories: (a) those that used the CTP parameter maps, and (b) those that used the native CTP time series to predict the AIS lesion.

The majority of the first group's methods used CNN-based models developed for medical image segmentation~\cite{crimi2018brainlesion,clerigues2019acute}. They have explored a wide range of deep learning techniques for predicting AIS lesions, including 2D~\cite{crimi2018brainlesion,soltanpour2021improvement}, 3D~\cite{abulnaga2018ischemic,crimi2018brainlesion}, and multi-modal predictive models~\cite{soltanpour2019ischemic,shi2021c,clerigues2019acute}. The U-Net~\cite{ronneberger2015u} and its extensions are the most commonly used model to perform the prediction task~\cite{ronneberger2015u}. Despite the large number of computations and complex design of these models, there are still some problems with them that prevent further improvements in the AIS lesion prediction task. A major limitation of these methods can be using predefined features, the CTP parameter maps, to predict AIS lesions~\cite{konstas2009theoretic}. Extracting these predefined maps from a CTP time series eliminates a lot of useful spatial and temporal information~\cite{mackay2003information}. Given these issues, deep neural networks, which tend to do better with raw data than they do with predefined high-level features, might not be the best tool for predicting AIS lesions using already manipulated CTP parameter maps~\cite{liang2017text}. 
Another problem is that these approaches rely on arterial input functions (AIFs) selection (\ie the time-concentration curve of one of the large feeding arteries of the brain~\cite{fieselmann2011deconvolution}) which is a highly error-prone process. Any underestimation or overestimation of the AIF selection will result in incorrect values for the perfusion parameter maps~\cite{robben2020prediction}.

Due to the shortcomings of using CTP parameter maps, the second group of approaches has increasingly explored the models that utilize native CTP time series for AIS lesions prediction. The majority of these methods have used the black-box deep learning models to predict AIS lesions from CTP time series images. 
For example, \cite{amador2021stroke} used multiple original CNN autoencoders for separately extracting features from each $3D$ volume snapshots of the CTP time series and then merging them to perform the final prediction using one decoder. 
In another study,~\cite{robben2020prediction} used different CNN encoders to extract features from the patients' clinical information (age, gender, etc.), the selected AIF spot, and the CTP time series to predict AIS lesions.
Moreover, there are numerous other methods that directly applied CNN autoencoder extensions, such as U-Net, to the native CTP time series~\cite{bertels2018contra,crimi2018brainlesion}. Typically, they perform a 3D prediction where the time dimension is regarded as the channels of the input layer of the CNN model.
Despite their ability to partially perform AIS prediction better than the `CTP maps' methods, these `CTP time series' methods are not yet adequate to be used in real-world applications. One of their flaws could be their ability to use the temporal information contained in the CTP time series, which is more crucial than spatial information to predict AIS lesions. 

Another concern that can affect both `CTP maps' and `CTP time series' group of methods is how they handle imbalanced classes in the AIS lesion prediction problem, \ie~AIS lesions usually include a small portion of the brain. In~\cite{bertels2018contra} the authors attempted to deal with imbalanced data problems by only using the affected hemisphere of the brain (AIS usually occurs in one hemisphere). In another study~\cite{pinheiro2018v}, they proposed to divide the $2D$ slices into smaller image batches, and excluded the batches not containing any lesion. The problem with these methods is that they learn to identify lesions only in images that are more likely to present them, and they are weak at identifying healthy brain regions. It should be noted that for a new image, where we do not know which hemisphere or slice does not contain the lesion to exclude, the model must be able to identify both healthy tissue and the lesion.



\begin{figure}[t]
	\begin{center}
		\includegraphics[width=1.05\textwidth]{{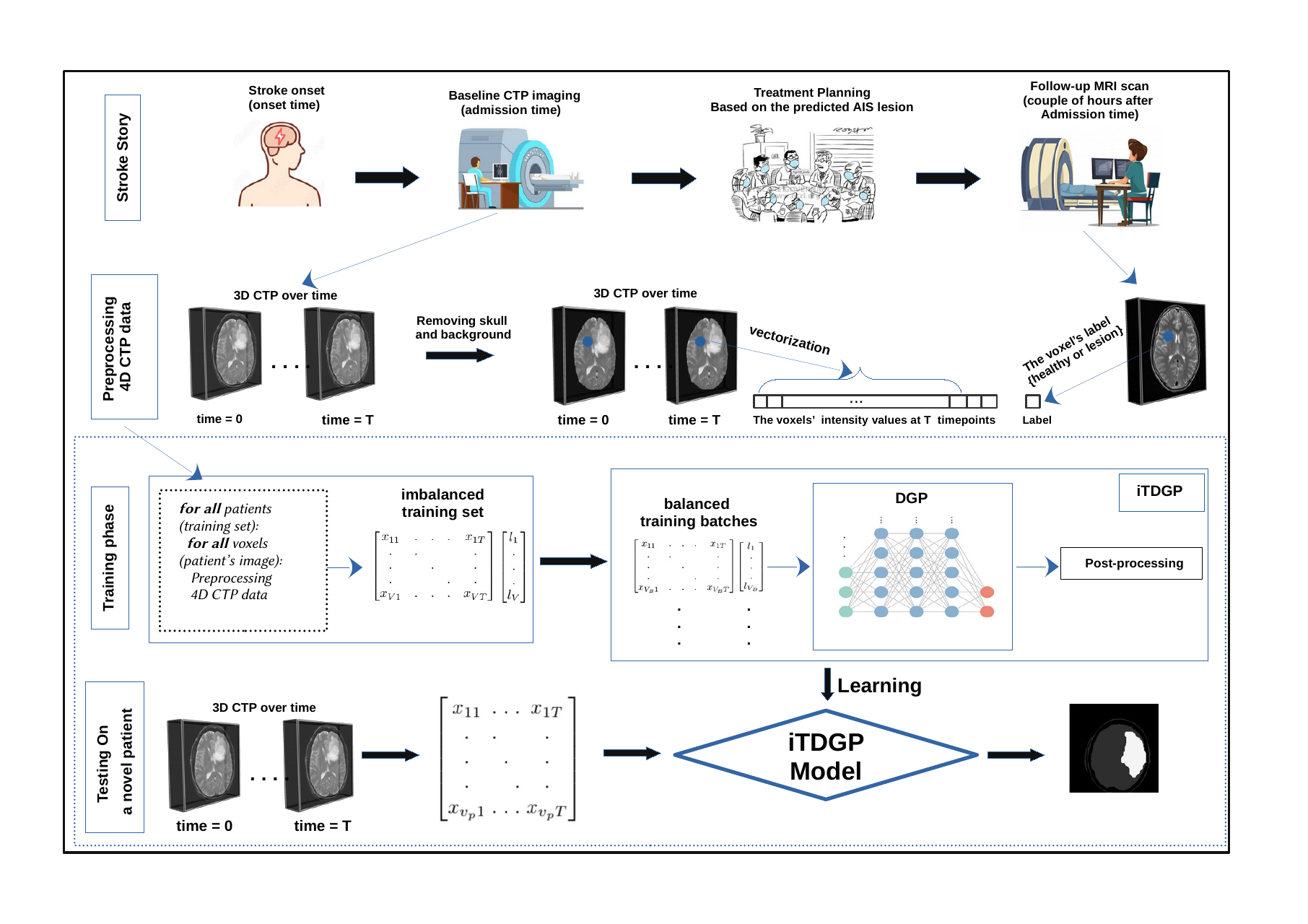}}
		\vskip -0.2in
	\caption{The first row (stroke story) summarizes the standard sequence of events associated with an emergency stroke event. Row 2 (preprocessing 4D CTP data) demonstrates the process of creating the temporal dataset. The 4D images are converted into a 2D matrix in which each row is a voxel, and the columns record their corresponding intensity values at the time points of CTP imaging.  In addition, using the follow-up MRI image, an expert determines the class label for each voxel (healthy or lesion). The third and fourth rows represent the training and testing phases, respectively.}
	\label{fig:framwork}
	\end{center}
 	\vskip -0.1in
\end{figure}

As it is presented in Figure \ref{fig:framwork}, this paper proposes a novel model that learns to predict AIS lesions based on native CTP time series: the imbalanced
Temporal Deep Gaussian Process (iTDGP), which employs Deep Gaussian Processes (DGP) to learn a probabilistic model that can effectively use temporal information to distinguish healthy voxels versus lesion ones. In contrast to other approaches, our proposed iTDGP does not require prior calibrations in the preprocessing phase --- \eg applying multi-subject anatomical alignments, \emph{etc}. 
Furthermore, iTDGP uses raw CTP time series volumes and does not require selecting arterial input functions (AIFs) (\ie the time-concentration curve of one of the large feeding arteries of the brain~\cite{fieselmann2011deconvolution}) which is a highly error-prone process.
We found that the iTDGP is better at identifying the AIS lesions by analyzing the entire CTP time series, rather than using previous methods that used spatial snapshots as channels of a CNN model, which is not an effective method to analyze the time series in a temporal manner~\cite{robben2020prediction,amador2021stroke}.
In addition, our proposed method is designed to perform robustly when the classes are imbalanced; note this is a major challenge for predicting AIS lesion, as there are many more healthy voxels than the lesion ones. 
During the learning process, iTDGP creates balanced batches, each of which includes all lesion voxels (small class), as well as the same number of instances that are randomly selected from healthy voxels (big class). 

We used two datasets to evaluate our proposed method and compare it with previously proposed methods. The first dataset was the ISLES 2018 training set that included the actual AIS lesion ground truths. Therefore, we were able to compare our method with the previous method by using the actual lesions' ground truth. The second dataset was our local dataset, the University of Alberta Hospital (UAH), which was collected for AIS future lesion prediction (after around 24 hours of stroke onset). 
Our comprehensive experiments, using the ISLES challenge 2018~\cite{cereda2016benchmarking} and UAH datasets, have shown that iTDGP is superior to the state-of-the-art AIS lesion predictors, obtaining a higher (cross-validation) Dice Similarity Coefficient (DSC).
Despite the promise of our proposed method in most experiments, iTDGP is a pixel-level classification method that may be slower than previous models that used end-to-end CNN models for AIS lesion prediction.

In the rest of the paper, Section \ref{sec:method} explains the proposed AIS lesion prediction method. Section \ref{sec:experiments} presents our experiments, achieved results, and discussions. Finally, Section \ref{sec:conclusion} presents the conclusions and suggests some future directions.


\section{Method}\label{sec:method}

This section describes iTDGP and how it can learn to predict both short- and long-term AIS lesions based on CTP time series. As depicted in (row 3 of) Figure~\ref{fig:framwork}, iTDGP learns a predictive model in three steps: generating balance batches, learning a deep Gaussian model, and smoothing the predicted area using spatial information. 
The first step addresses the imbalanced sample size challenge (mentioned in the introduction), as iTDGP first generates a set of balanced batches from the AIS dataset --- where each batch includes all instances from the small class (lesion voxels) and randomly draws the same number of instances (equal to small class size) from the big class. 
Next, iTDGP learns the model parameters from these balanced batches. Lastly, iTDGP uses a heavy-tailed Gaussian smoothing procedure to enhance the robustness of predicted regions.

It should be mentioned that the methods, experiments, as well as results presented in this paper, have been verified by relevant physicians in the neurology division of the University of Alberta hospital based on the relevant guidelines. Additionally, both datasets we used --- the public ISLES 2018 dataset~\cite{cereda2016benchmarking} and the local UAH dataset --- were anonymous and collected with consent from all subjects or their legal guardians. 
The UAH dataset contained $55$ patients collected as part of an ongoing prospective stroke study. The study was approved by the Health Ethics Committee of the University of Alberta (Pro00066577). Written informed consent was received from all study participants. The dataset contained $55$ patients who had the follow-up MRIs within 24(±6) hours of their baseline CTP imaging.

Let $T$ be the number of time points, and $V$ be the number of voxels for training set. Note that the preprocessing procedure includes the time synchronization step~\cite{robben2020prediction}, which selects the same number of time points for each patients --- here, we used $T = 20$ for ISLES and $T=40$ for UAH. The preprocessed CTP time series for the training phase use the input matrix $\mathbf{X} \in \mathbb{R}^{V \times T}$ and the label vector $\mathbf{y} \in \left \{ 0,1 \right \}^{V}$. Let $V_L$ be the number of voxels in the large class (selected based on training label $\mathbf{y}$) and $V_S$ be the number of voxels in the small class. Our proposed imbalanced approach then generates $B=[\frac{V}{V_S}]$ balanced batches for each epoch, in which each batch consists of all voxels of the small class and the $V_S$ samples of the large class --- \ie{randomly sampled without replacement}. Here, $[\cdot]$ denotes the integer part function. We let $E$ be the number of epochs in the training phase, and $V_B = 2V_S$ be the number of voxels in each batch. We respectively define $\mathbf{X}^{(s)} \in \mathbb{R}^{T\times V_B}$ and $\mathbf{y}^{(s)} \in \left\{ 0,1 \right\}^{V_B}$ for $s=1, 2,\dots, E\times B$ as features and labels for $s\text{-}th$ training iteration. In next section, we explain how iTDGP uses $\mathbf{X}^{(s)}$ and $\mathbf{y}^{(s)}$ in each iteration to learn a predictive model.

\subsection{imbalanced Temporal Deep Gaussian Process (iTDGP)}
Our proposed iTDGP is a fully-connected deep neural network that is comprised of $H$ Gaussian Processes (GPs) layers. The proposed iTDGP offers the following properties:

\begin{itemize}
    \item iTDGP can map each hidden layer to the next in a more expressive and data-driven manner than a pre-fixed sigmoid non-linear function, which is used in standard parametric deep learning algorithms~\cite{kumar2018deep}.
    
    \item iTDGP uses an Automatic Relevance Determination (ARD) kernel that can automatically determine its layers dimensionality --- \ie we do not need to manually assign the number of neurons~\cite{kumar2018deep,murphy2012machine}.

    \item  iTDGP is robust against overfitting small data due to its averaging and stochasticity characteristics of GPs~\cite{lawrence2005probabilistic}.
\end{itemize}



Let $\mathbf{x}_{v}^{(s)} \in \mathbb{R}^{T}$, $\mathbf{X}^{(s)} = \{ \mathbf{x}_{v}^{(s)} \}_{v=1}^{V_B} $ and $y_{v}^{(s)} \in \{0, 1\}$, $\mathbf{y}^{(s)} = \{ y_{v}^{(s)} \}_{v=1}^{V_B}$ for~$s=1, 2,\dots, (E\times B)$~be respectively the $v\text{-}th$ instance and its corresponding label from $s\text{-}th$ training iteration. Letting $H$ be the number of iTDGP network layers, it uses a composition of functions $f(\mathbf{x}_{v}^{(s)}) = f^{(H)} \circ (f^{(H-1)}...\circ(f^{(1)}(\mathbf{x}_{v}^{(s)})))$ to map each data point $\mathbf{x}_{v}^{(s)}$, to its associated output $y_{v}^{(s)}$. Here, function $f^{(h)}$ for the intermediate layer $h=1\dots H-1$ includes $T^{(h)}$ GP functions (\ie $f^{(h)}= \left \{ f_{j}^{(h)} \right \}_{j=1}^{T^{(h)}}$) that maps representations of $h\text{-}th$ to $(h+1)\text{-}th$ layer. Further, $f_{j}^{(h)}$ in layer $h$ provides $j\text{-}th$ representation by using independent GP priors --- \ie $f_j^{(h)}( \mathbf{x}_{v}^{(s)} ) \sim\mathcal{GP}(m_{j}^{(h)}(\mathbf{x}_{v}^{(s)}), k^{(h)}(\mathbf{x}_{v}^{(s)}, \mathbf{x}_{\hat{v}}^{(s)}))$, where $v,\hat{v} \in \{ 1\dots {V}_{B} \}$, $m_{j}^{(h)}(\mathbf{x}_{v}^{(s)})$ is mean function and $k^{(h)}(\mathbf{x}_{v}^{(s)}, \mathbf{x}_{\hat{v}}^{(s)})$ denotes the covariance function for $j\text{-}th$ representation in layer $h$ \cite{murphy2012machine}. We let ${F}^{(1)}_{v,j}=f_j^{(1)}(\mathbf{x}^{(s)}_{v})$ be the representation for $j\text{-}th$ function of the $1\text{-}st$ layer that produced using $\mathbf{x}^{(s)}_{v}$ input data, and $\mathbf{F}_{v}^{(h)} = \{F_{v,j}^{(h)} \}_{j=1}^{T^{(h)}}$ denote all representations for $v\text{-}th$ instance in layer $h$.
We then define $F_{v,j}^{(h)}=f_{j}^{(h)}(\mathbf{F}_{v}^{(h-1)})$ as the $j\text{-}th$ representation for $h=2\dots H-1$ intermediate layers. Note the last layer $\mathbf{F}^{(H)} = f^{(H)}(\cdot)$ consists of one neuron that uses a soft-max function to perform the binary classification task. We also let $\mathbf{f}^{(h)}_{j} = \{ {F}^{(h)}_{v,j} \}_{v=1}^{V_B}$ denote the $j\text{-}th$ representation at layer $h$ that computed by using all input data, and $\mathbf{F} = \{ \mathbf{F}^{(h)} \}_{h=1}^{H}$ be all representations in all layers. 
The kernel function corresponding to GPs at layer $h$ is expressed as~\cite{kumar2018deep}:
\begin{equation}\label{eq:kernel}
    k^{(h)}(\mathbf{F}_{v}^{(h)},\mathbf{F}_{\hat{v}}^{(h)}) = (\sigma^{(h)}_{f})^{2}exp\Big(-\frac{1}{2\omega^{(h)}} \left \| \mathbf{F}_{v}^{(h)}-\mathbf{F}_{\hat{v}}^{(h)} \right \|^2\Big), 
\end{equation}
for each $v,\hat{v} \in \{ 1\dots {V}_{B} \}$. Using the weighted covariance function \eqref{eq:kernel}, it is possible to assign different weights to different latent dimensions. This can be used in a Bayesian training framework to turn off irrelevant dimensions by reducing their weights to zero. By applying this procedure, complex models will be automatically structured. 
However, since the covariance function introduces non-linearity to this model, it is difficult to apply Bayesian methods to it~\cite{kumar2018deep}. In order to cope with this issue, we can define a Bayesian training procedure analytically based on recent nonstandard variational inference methods~\cite{williams2006gaussian}.

\subsection{Bayesian Training to Optimize iTDGP}\label{sec:Bayesian Training}
We follow~\cite{kumar2018deep,williams2006gaussian} for optimizing the parameters of our proposed iTDGP model. There are a few parameters and variables used in this section that are defined in the Method Section of the main paper.
Based on variational sparse Gaussian Process (GP) approximation~\cite{williams2006gaussian}, each layer has an induced variable set $\mathbf{U} = \left \{ \mathbf{U}^{(h)} \right \}_{h=1}^{H-1}$, where each $\mathbf{U}^{(h)} = \{\mathbf{u}_{j}^{(h)} \}_{j=1}^{T^{(h)}}$ --- $H$ is the number of layers in the network, and $T^{(h)}$ is the number units in $h\text{-}th$ layer. This is a set of function values over $M$ inducing points $\mathbf{Z}^{(h)}$ associated with layer $h$,
$\mathbf{Z}^{(h)} = \left \{ \mathbf{z}_{i}^{(h)} \right \}_{i=1}^{M}$, 
where each $\mathbf{u}_{j}^{(h)} $ is the inducing variables associated with the $j\text{-}th$ representation at the $h\text{-}th$ layer. 
For simplicity, we considered a fix number of inducing points $M$, for all layers which is followed by a joint GP prior over inducing points as well as latent function values. The joint distribution $p(\mathbf{y}, \mathbf{F}, \mathbf{U} )$ is computed by:
\begin{equation}
\begin{gathered}
    \prod_{v=1}^{V_B}p(y_v\mid \mathbf{F}_{v}^{(H)})\prod_{h=1}^{H}\prod_{j=1}^{T^{(h)}}p(\mathbf{f}_{j}^{(h)}\mid\mathbf{u}_{j}^{(h)},\mathbf{F}^{(h-1)},\mathbf{Z}^{(h)})\: p(\mathbf{u}_{j}^{(h)}\mid \mathbf{Z}^{(h)}).
\end{gathered}
\end{equation}
We respectively define $\mathbf{X}^{(s)} \in \mathbb{R}^{T\times V_B}$ and $\mathbf{y}^{(s)} \in \left\{ 0,1 \right\}^{V_B}$ for $s=1, 2,\dots, E\times B$ as features and labels for $s\text{-}th$ training iteration, where $E$ the number of epochs in the training phase, $B$ is the number of balanced batches, and ${V_B}$ is the number of voxels in each created balanced batches. 
In this case, a deep GP prior is applied recursively over all latent spaces with $F^0=\mathbf{X}^{(s)}$ and soft-max likelihood is utilized for classification. 

We let $\mathbf{K}_{XY}$ be a covariance matrix over the general matrices $X$ and $Y$ \cite{kumar2018deep}. 
We then define the mean and covariance functions as follows:
\begin{equation}
\begin{gathered}
\text{mean}(\mathbf{f}_{j}^{(h)}) = m_{j}^{(h)}(\mathbf{F}^{(h-1)})+\mathbf{K}_{\mathbf{F}^{(h-1)}\mathbf{Z}^{(h)}}^{(h)} (\mathbf{K}_{\mathbf{Z}^{(h)}\mathbf{Z}^{(h)}}^{(h)})^{-1} (\mathbf{u}_{j}^{(h)}-m_{j}^{(h)}(\mathbf{Z}^{(h)})),
\end{gathered}
\end{equation}
\begin{equation}
\begin{gathered}
\text{cov}(\mathbf{f}_{j}^{(h)})=\mathbf{K}_{\mathbf{F}^{(h-1)}\mathbf{F}^{(h-1)}}^{(h)}-\mathbf{K}_{\mathbf{F}^{(h-1)}\mathbf{Z}^{(h)}}^{(h)}(\mathbf{K}_{\mathbf{Z}^{(h)}\mathbf{Z}^{(h)}}^{(h)})^{-1}(\mathbf{K}_{\mathbf{F}^{(h-1)}\mathbf{Z}^{(h)}}^{(h)})^{\top }.
\end{gathered}
\end{equation}
We also let $\mathcal{N}$ operator as the normal distribution. The conditional probability for $\mathbf{f}_{j}^{(h)}$ is then defined as follows:
\begin{equation}
    p(\mathbf{f}_{j}^{(h)}\mid\mathbf{u}_{j}^{(h)}, \mathbf{F}^{(h-1)},\mathbf{Z}^{(h)})= \mathcal{N}(\mathbf{f}_{j}^{(h)};\text{mean}(\mathbf{f}_{j}^{(h)}), \text{cov}(\mathbf{f}_{j}^{(h)})).
\end{equation}
Since obtaining the marginal prior over $\left \{ \mathbf{F}^{(h)} \right \}_{h=2}^{H}$ is intractable, the posterior distribution $p(\mathbf{F},\mathbf{U}\mid \mathbf{y})$ and marginal likelihood $p(\mathbf{y})$ cannot be computed in closed form~\cite{kumar2018deep}.
To cope with this problem, we use variational inference approach proposed in~\cite{damianou2013deep}, where the variational posteriors are computed as follows:
\begin{equation}
    q(\mathbf{F},\mathbf{U}) = \prod_{h=1}^{H}\prod_{j=1}^{T^{(h)}}p(\mathbf{f}_{j}^{(h)}\mid\mathbf{u}_{j}^{(h)},\mathbf{F}^{(h-1)},\mathbf{Z}^{(h)})\:q(\mathbf{u}_{j}^{(h)}),
\end{equation}
where $q(\mathbf{u}_{j}^{(h)})= \mathcal{N}(\mathbf{u}_{j}^{(h)}; \mathbf{m}_{j}^{(h)},\mathbf{S}_{j}^{(h)})$, $\mathbf{m}^{(h)} = \{ \mathbf{m}_{j}^{(h)} \}_{j=1}^{T^{(h)}}$ is a vector formed by concatenating the vectors $\mathbf{m}_{j}^{(h)}$, and $\mathbf{S}^{(h)}$ is the block diagonal covariance matrix constructed by $\mathbf{S}_{j}^{(h)}$. 
This method can be extended to multiple layers by using the following methodology:
\begin{equation}\label{eq:extend}
\begin{gathered}
    L(\left \{ \mathbf{m}^{(h)}, \mathbf{S}^{(h)} \right \}_{h=1}^{H})=\sum_{v=1}^{V_B}\mathbb{E}_{q(\mathbf{F}_{v}^{(H)})}\left [\log p(y_v\mid \mathbf{F}_{v}^{(H)}) \right ]- \sum_{v=1}^{H}KL[q(\mathbf{U}^{(h)})\mid\mid p(\mathbf{U}^{(h)})],
\end{gathered}
\end{equation}
where KL is Kullback–Leibler divergence~\cite{hershey2007approximating}, and $E_x[\cdot]$ denotes the expected value function with respect to $x$ distribution.
In addition, over all the data points, the marginal distribution of the functions values for layer $H$ can be computed as:
\begin{equation} \label{eq:marginaldistribution}
\begin{gathered}
    q(\mathbf{F}^{(H)}\mid \left \{ \mathbf{Z}^{(h)},\mathbf{m}^{(h)}, \mathbf{S}^{(h)} \right \}_{h=1}^{H})=
    \int_{\mathbf{F}^{(1)},...,\mathbf{F}^{(H-1)}}\prod_{h=1}^{H}q(\mathbf{F}^{(h)}\mid \mathbf{F}^{(h-1)},\mathbf{Z}^{(h)},\mathbf{m}^{(h)}, \mathbf{S}^{(h)})d\mathbf{F}^{(1)},...,d\mathbf{F}^{(H-1)},
\end{gathered}
\end{equation}
To obtain the conditional distribution in equation \eqref{eq:marginaldistribution} we used:
\begin{equation}\label{eq:pareto mle2}
  \begin{gathered}
    q(\mathbf{F}^{(h)}\mid \mathbf{F}^{(h-1)},\mathbf{Z}^{(h)},\mathbf{m}^{(h)}, \mathbf
    {S}^{(h)})=
    \prod_{j=1}^{T^{(h)}}\int_{\mathbf{u}_{j}^{(h)}}p(\mathbf{f}_{j}^{(h)}\mid \mathbf{u}_{j}^{(h)},\mathbf{F}^{(h-1)},\mathbf{Z}^{(h)})q(\mathbf{u}_{j}^{(h)})d\mathbf{u}_{j}^{(h)} 
    = \prod_{j=1}^{{h}}\mathcal{N}(\mathbf{f}_{j}^{{(h)}}; \mathbf{\tilde{m}}_{j}^{{(h)}} ,\mathbf{\tilde{V}}_{j}^{{(h)}})
 \end{gathered}
\end{equation}
where we have
\begin{equation}
\begin{gathered}
\mathbf{\tilde{m}}_{j}^{(h)} = m_{j}^{(h)}(\mathbf{F}^{(h-1)})+\mathbf{K}_{F^{(h-1)}\mathbf{Z}^{h}}^{(h)}(\mathbf{K}_{\mathbf{Z}^{(h)}\mathbf{Z}^{(h)}}^{(h)})^{-h}(\mathbf{m}_{j}^{(h)}- m_{j}^{(h)}(\mathbf{Z}^{(h)})),
\end{gathered}
\end{equation}
\begin{equation}
\begin{gathered}
\mathbf{\tilde{V}}_{j}^{(h)} =\mathbf{K}_{\mathbf{F}^{(h-1)}\mathbf{F}^{(h-1)}}^{(h)}-\mathbf{K}_{\mathbf{F}^{(h-1)}\mathbf{Z}^{(h)}}^{(h)}(\mathbf{K}_{\mathbf{Z}^{(h)}\mathbf{Z}^{(h)}}^{(h)})^{-1}(\mathbf{K}_{\mathbf{Z}^{(h)}\mathbf{Z}^{(h)}}^{(h)}-\mathbf{S}_{j}^{(h)})(\mathbf{K}_{\mathbf{Z}^{(h)}\mathbf{Z}^{(h)}}^{(h)})^{-1}(\mathbf{K}_{F^{(h-1)}\mathbf{Z}^{(h)}}^{(h)})^{\top}.
\end{gathered}
\end{equation}
The non-liner stochastic term~$\left \{ \mathbf{F}^{(h-1)} \right \}_{h=2}^{H}$ within the conditional distributions \\$\left \{q( \mathbf{F}^{(h)} \mid \mathbf{F}^{(h-1)}, \mathbf{m}^{(h)}, \mathbf{Z}^{(h)}, \mathbf{S}^{(h)} \right \}_{h=2}^{H-1}$, makes the marginal distribution in equation~\eqref{eq:marginaldistribution} intractable.
This intractability results in the expected log likelihood in equation~\eqref{eq:extend} to be intractable even for Gaussian likelihood. We used the Monte Carlo sampling approximation method proposed in~\cite{salimbeni2017doubly}. In this approach, the marginal variational posterior over function values in the final layer for $v\text{-}th$ data point --- \ie $q(\mathbf{F}_{v}^{(H)})$ --- depends only on the $v\text{-}th$ marginal so fall the previous layers. Each $\mathbf{F}_{v}^{(h)}$ is sampled from $q(\mathbf{F}_{v}^{(h)}\mid \mathbf{F}_{v}^{(h-1)}, \mathbf{Z}^{(h)}, \mathbf{m}^{(h)}, \mathbf{S}^{(h)})=\mathcal{N}(\mathbf{F}_{v}^{(h)}; \mathbf{\tilde{m}}^{(h)}[v], \tilde{\mathbf{V}}^{(h)}[v] )$ where $\mathbf{\tilde{m}}^{(h)}[v]$ ($T^{(h)}$-dimensional vector) and $\tilde{\mathbf{V}}^{(h)}[v]$ are respectively the $T^{(h)}$-dimensional mean and the $T^{(h)} \times T^{(h)}$ diagonal covariance matrix of the $v\text{-}th$ data point
over representations in layer $h$ , which shows how it depends on $\mathbf{F}_{v}^{(h-1)}$.
We use the `reparametarization trick', where the sampling can be written as \cite{kumar2018deep,salimbeni2017doubly}:
\begin{equation}
\begin{gathered}
      \mathbf{F}_{v}^{(h)}= \mathbf{\tilde{m}}^{(h)}[v]+ \mathbf{\epsilon} ^{(h)} \odot \tilde{\mathbf{V}}^{(h)}[v]^{\frac{1}{2}}, \qquad  \mathbf{\epsilon}^{(h)} \sim \mathcal{N}( 0,\mathbb{I}_{T^{(h)}})
\end{gathered}
\end{equation}

Using a mini-batch of data, one can compute gradients on the lower bound, and the parameters are updated based on the sum of the data points. This allows one to maximize the variational lower bound by using stochastic gradient techniques. Combining the stochasticity of gradient computation with Monte Carlo sampling to compute variational lower bounds leads to a doubly stochastic inference method for deep GPs.



\subsection{Postprocessing}
After classifying each voxel, iTDGP converts the labelled voxels to the image containing the predicted AIS lesion. Let $\mathbf{G}$ denote a smoothing kernel with log-normal distribution~\cite{wilcox2003applying} and $\mathbf{X}$ be the input image. Then the learned neural network within iTDGP model, $f(\cdot) = f^{(H)} \circ (f^{(H-1)}...\circ(f^{(1)}(\cdot)))$, will produce the predicted labels for each voxel of $\mathbf{X}$ --- \ie the segmented image, $\mathbf{Y} = f(\mathbf{X})$. The postprocessed prediction is $\mathbf{\tilde{Y}} = \mathbf{G}*\mathbf{Y}$, where $*$ is the convolution operator. Based on our experimental results, we found that this postprocessing creates a more homogeneous predicted lesion by removing tiny misclassified regions. 
Since AIS lesions usually appear as homogeneous masses~\cite{van2007acute},
the postprocessing step enable us to 
improve prediction accuracy.

\section{Experiments and Results}\label{sec:experiments}
We evaluated our iTDGP model for AIS lesion prediction using two datasets. (1) To predict short-term AIS lesions, we used the public dataset from \textbf{ISLES challenge 2018}~\cite{cereda2016benchmarking}, which includes $94$ scans from $63$ patients (taking 2 slabs from some patient to illustrate the entire AIS lesion)
collected from two imaging centers. 
Here, each patient had an Magnetic Resonance Imaging (MRI) within $3$ hours of their baseline CTP imaging; 
Using the short-term follow-up MRI scans, the experts obtained the ground-truths of the future AIS lesions.
For each patient, there are different numbers of brain image slices, from $2$ to $16$, but all the slices (both the baseline CTP and follow-up MRI) are the same size $256\times 256$.

(2) The \textbf{UAH} (a private) dataset contained $55$ treated patients who had the follow-up MRIs within $24(\pm 6)$ hours of their baseline CTP imaging. For each patient, we labeled ground-truth future AIS lesions based on their follow-up MRI. Here both the baseline CTP and follow-up MRIs have slice size of $512\times 512$. To be consistent, we downsampled the UAH dataset images (both CTP and MRI) by a $2\times2$ kernel so that they are the same image size as ISLES dataset images. 

Note that, in both datasets, clinical information was not included in our analysis. In this study, we focused just on using imaging information and comparing our results with previous works that only used the same information. In the future, we plan to investigate the impact of including other clinical information, including treatment type, time intervals, gender, age, etc, on AIS prediction accuracy.

To fairly compare the obtained results, we evaluated all the models using one-patient-out cross-validation: training on all but one patient, then testing on that held-out patient. In accordance with previous AIS prediction studies, we have reported Dice Similarity Coefficient (DSC), Jaccard, precision and recall scores to show the performance of the proposed prediction model.
We implemented our proposed model with PyTorch, optimized with the Adam algorithm~\cite{kingma2014adam} on NVIDIA GeForce GTX 980M GPU. Based on the poly schedule, ${\alpha_{iteration} = \alpha_{iteration-1} \times(1 - \frac{iteraion}{total\,iterations})}$, the learning rate $\alpha$ is initially set to $0.01$ and decays over iterations.

\subsection{Preprocessing}

The first step of preprocessing was to remove the skull and background from the CTP images.
Then, in order to enhance the Signal-to-Noise Ratio (SNR) of the CTP images, which is critically low, we applied an in-slice $2D$ spatial Gaussian filter to reduce noise and smooth the images. 
We then converted the $4D$ CTP time series images of each patient $p \in \left \{1, \dots, P  \right \} $ (where P is the number of the patients in the dataset) into a $2D$ temporal matrix of $\mathbf{X}_{p} \in R^{v_{p} \times t_{p}} $, and labeled the voxels' label in  $\mathbf{y}_{p}\in \left \{ 0,1 \right \}^{v_{p}} $, where $v_{p}$ is the number of brain voxels of patient number $p$, and $t_{p}$ is the number of time points for that patient. 
Finally, to make the temporal matrices comparable across the patients, we separately normalized them by:~$normalized(\mathbf{X}_p) = (\mathbf{X}_p - mean(\mathbf{X}_p))/std(\mathbf{X}_p)$.

Because of the differences in brain volume sizes and imaging time length, the values of  the values of $v_p$ are different, as are the values of $t_p$. 
To equalize the number of time points (that are considered as features in our method) for all the patients, we found the patient with the smallest $t_p$ and named it $T$ (here, $T = 20 $ for ISLES and $40$ for UAH). Then, for every other patient, we removed the first and last  $(t_p - T)/2$  time point.
Finally, we concatenated all of these $2D$ temporally equalized matrices from all the patients in to the matrix $\mathbf{X}\in R^{V \times T} $, where \textit{V} is the total number of voxels in the whole dataset and \textit{T} is the equalized number of time points. The labels of the voxels are also stored in $\mathbf{y}\in \left \{ 0,1 \right \}^{V} $.

\begin{figure*}[t]
	\begin{center}
		\begin{minipage}{0.4\linewidth}\includegraphics[width=1\textwidth]{{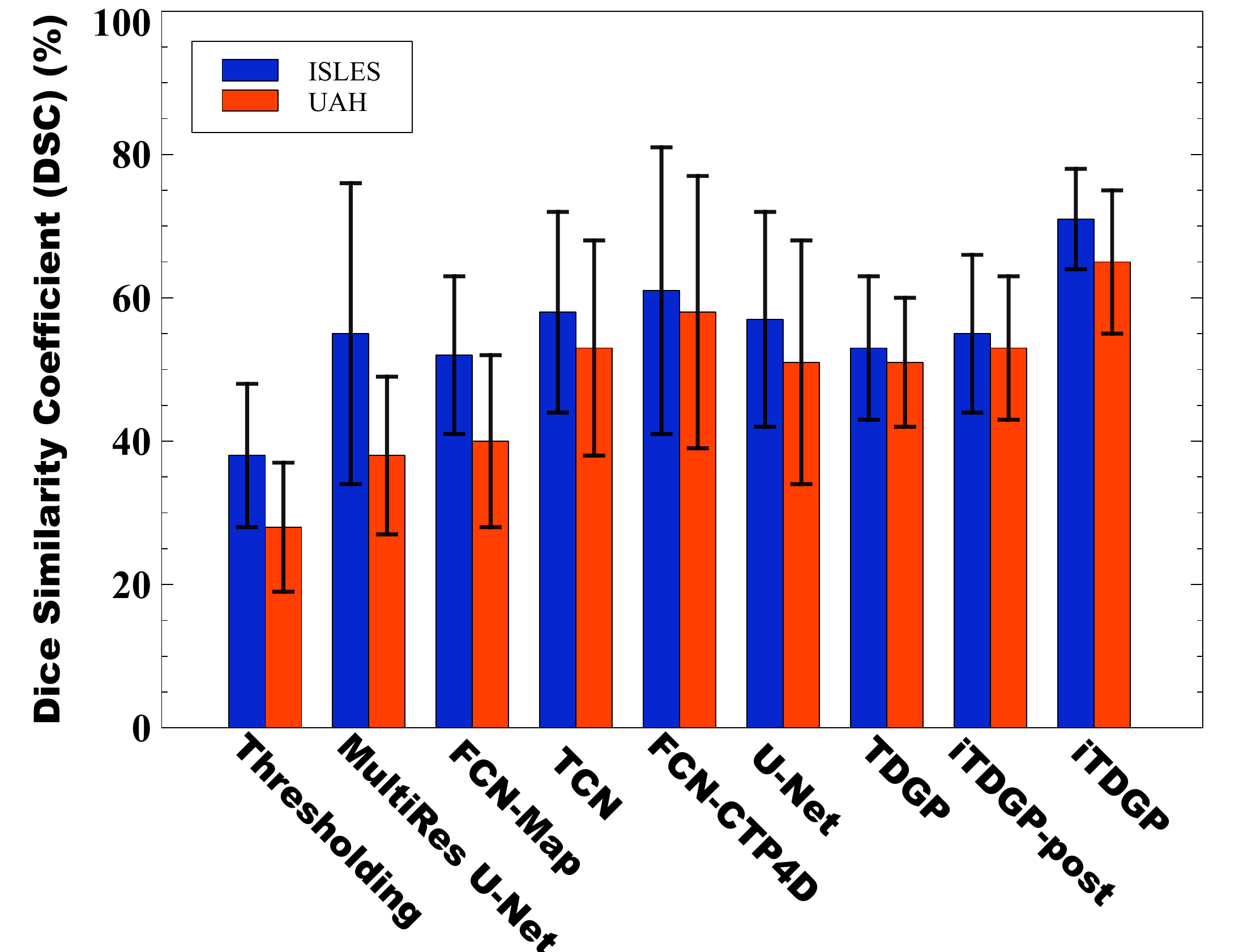}}
		\end{minipage}
		\begin{minipage}{0.4\linewidth}\includegraphics[width=1\textwidth]{{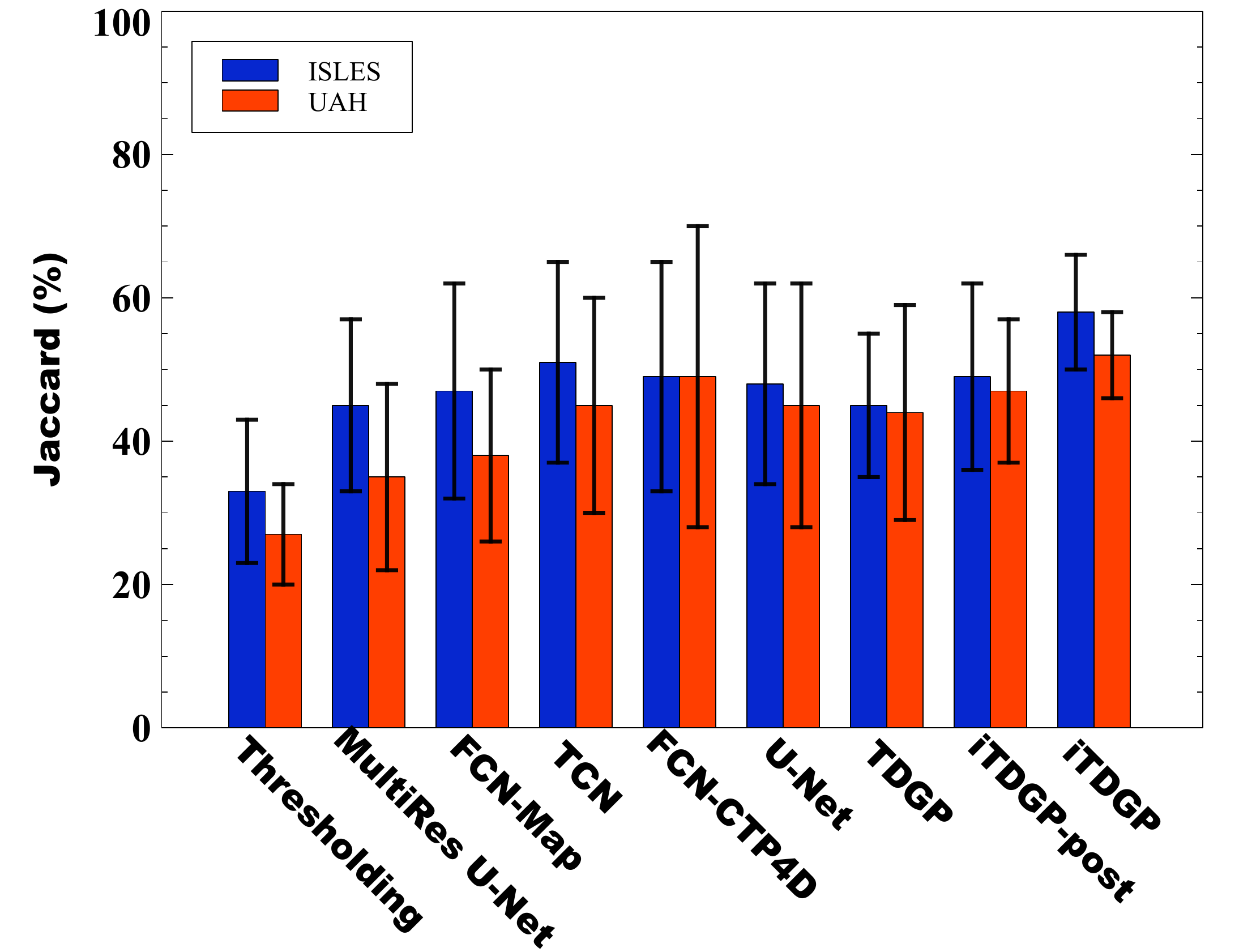}}
		\end{minipage}
		\begin{minipage}{0.4\linewidth}\includegraphics[width=1\textwidth]{{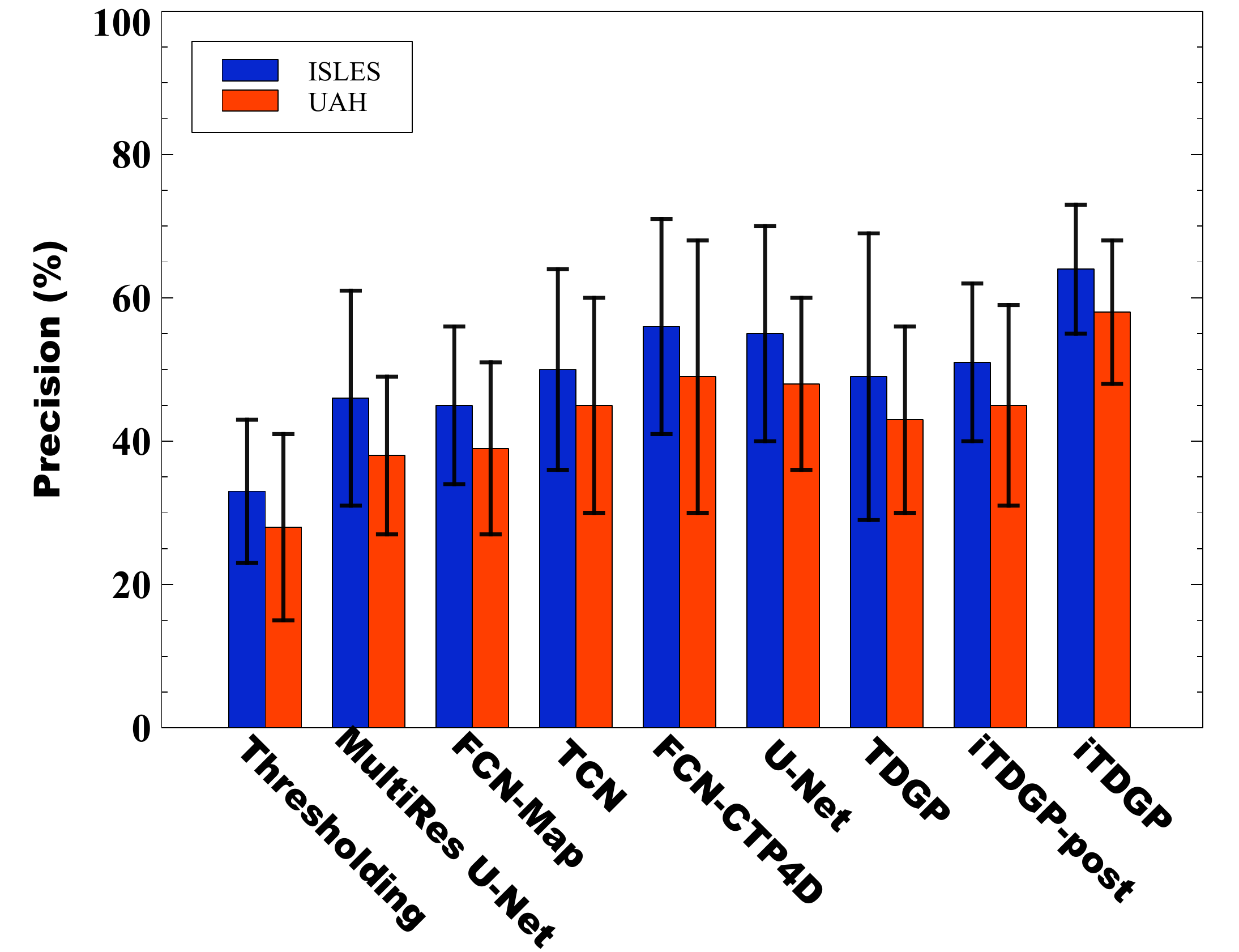}}
		\end{minipage}
		\begin{minipage}{0.4\linewidth}\includegraphics[width=1\textwidth]{{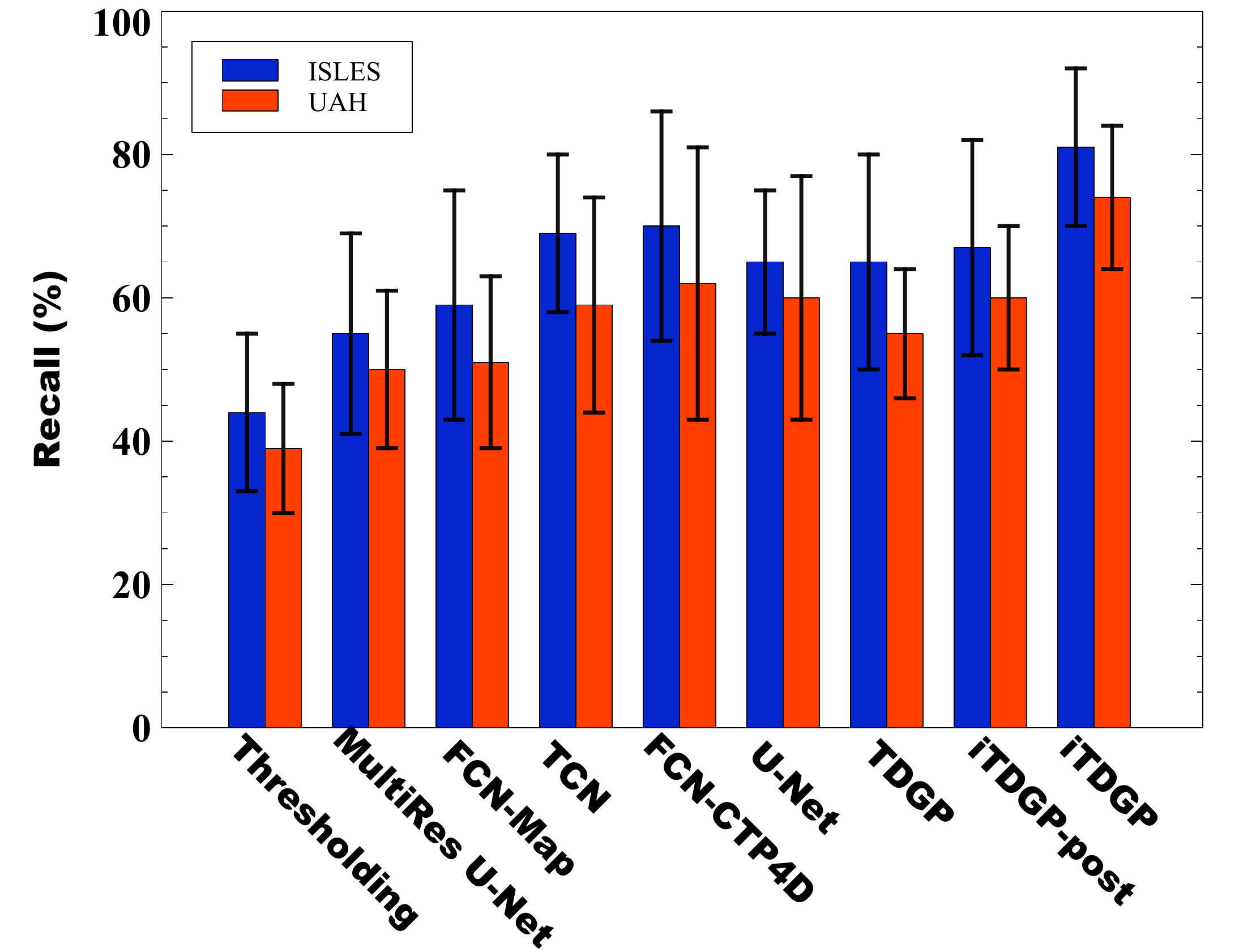}}
		\end{minipage}
	\end{center}
		\vskip -0.2in
	\caption{AIS lesion scores (DSC, Jaccard, Precision, Recall) obtained by $9$ different predictive models: Thresholding (baseline), Multires U-Net~\cite{soltanpour2021improvement}, FCN-Map~\cite{clerigues2019acute}, TCN~\cite{amador2021stroke}, FCN-CTP4D (metadata excluded)~\cite{robben2020prediction}, U-Net~ \cite{bertels2018contra}, TDGP (iTDGP without postprocessing and balanced batches), iTDGP-post (iTDGP without postprocessing), and iTDGP. Note that larger numbers are good, for all 4 measures. Also, the bars show 1 std dev. }
	\label{fig:Performance}
			\vskip -0.1in
\end{figure*}

\subsection{Prediction Scores}
Figure~\ref{fig:Performance} shows the results of some recently proposed AIS lesion predictors, as well as our proposed model. 
The first three methods from the left (Thresholding, MultiRes U-Net~\cite{soltanpour2021improvement}, and FCN-Map~\cite{clerigues2019acute}) are `CTP parameter map' models, which use only the CTP parameter maps to predict AIS lesion. The next 3 methods --- \ie TCN~\cite{amador2021stroke}, FCN-CTP4D (metadata excluded)~\cite{robben2020prediction}, and U-Net~\cite{bertels2018contra} --- used native CTP time series for the lesion prediction problem. 
The rightmost three methods in figure~\ref{fig:Performance} are our proposed methods, \ie  `TDGP' uses neither balanced batches nor postprocessing, `iTDGP-post' used balanced batches without postprocessing step, and `iTDGP' is the complete model.
One can see that the models that used the CTP time series are superior to the `CTP parameter map' models. In addition, our proposed prediction method performs the prediction task more accurately than the other `CTP time series' methods. Note each of the 4 plots in figure \ref{fig:Performance} is comparing our iTDGP and $\chi$, for 6 different $\chi\in \{ \text{baseline, Multires U-Net~\cite{soltanpour2021improvement}, FCN-Map~\cite{clerigues2019acute}, TCN~\cite{amador2021stroke}, FCN-CTP4D~\cite{robben2020prediction}, U-Net~ \cite{bertels2018contra}} \}$, and using two datasets, ISLES and AHS. Then for a total of $4\times 6\times 2=48$ comparisons, 2-sided t-test found $p<0.05$ in all 48 cases.

Recall that the ISLES dataset is collected to predict short-term future AIS lesions, while the UAH dataset is collected to predict long-term future AIS lesions. In Figure~\ref{fig:Performance}, the blue blocks, which are related to short-term AIS lesion prediction, have a lower increment rate from left to right --- `CTP maps' to `CTP time series' models --- than the red blocks, which are for long-term prediction. This suggests that the temporal information from the CTP time series might be more relevant for long-term AIS lesions prediction than for short-term predictions. 

It should be noted that our proposed iTDGP is actually a voxel-level classifier model. Consequently, this model can be slower - but not so slow as to affect the clinical requirements of the application - than others using deep neural networks. However, in this study, we focused on improving the accuracy of the model rather than its speed.

\begin{figure}[t]
\begin{center}
	\begin{minipage}{0.8\linewidth}
		\centering
		    \includegraphics[width=0.99\textwidth]{{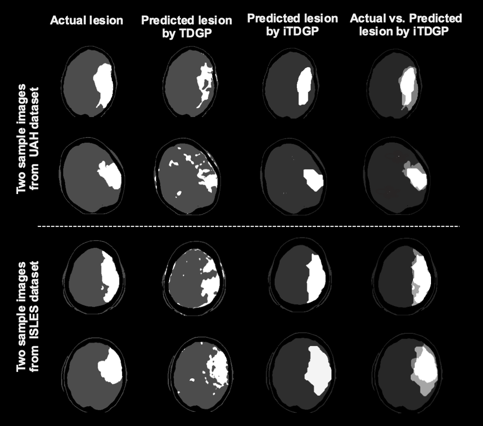}}
		\caption{Visualization of AIS lesion predicted by the proposed iTDGP, versus TDGP, which does not include the imbalanced and post-processing extensions.}
		\label{fig:Visualize}
	\end{minipage}
  	\hspace{0.1cm}
	
\end{center}
\vskip -0.15in        
\end{figure}

\begin{figure}[t]
\begin{center}
	
  	\hspace{0.1cm}
	\begin{minipage}{0.9\linewidth} 
		\centering
		\begin{minipage}[b]{0.6\linewidth}
        	\includegraphics[width=0.98\textwidth]{{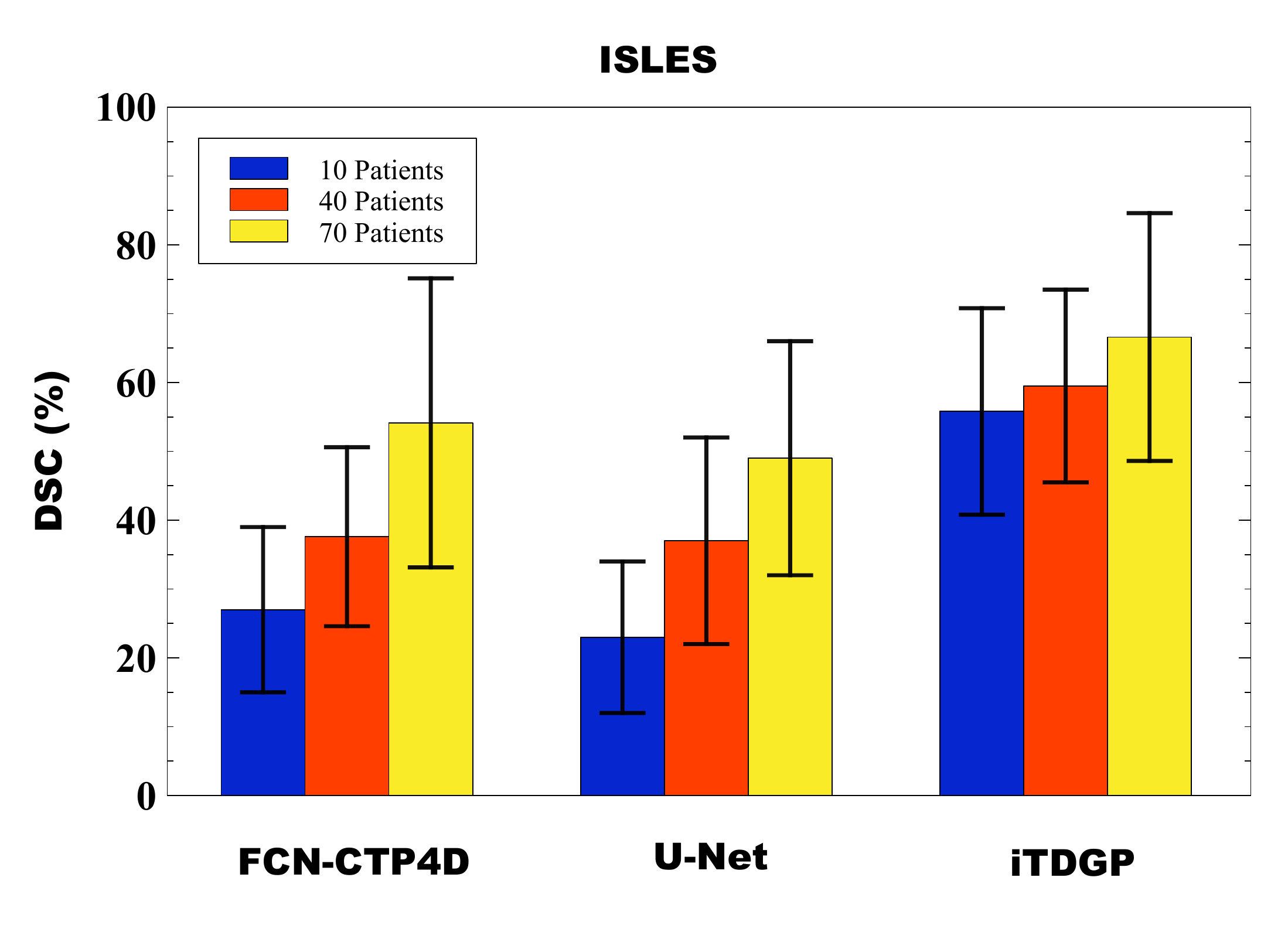}}
        \end{minipage}
        \begin{minipage}[b]{0.6\linewidth}
	        \includegraphics[width=0.98\textwidth]{{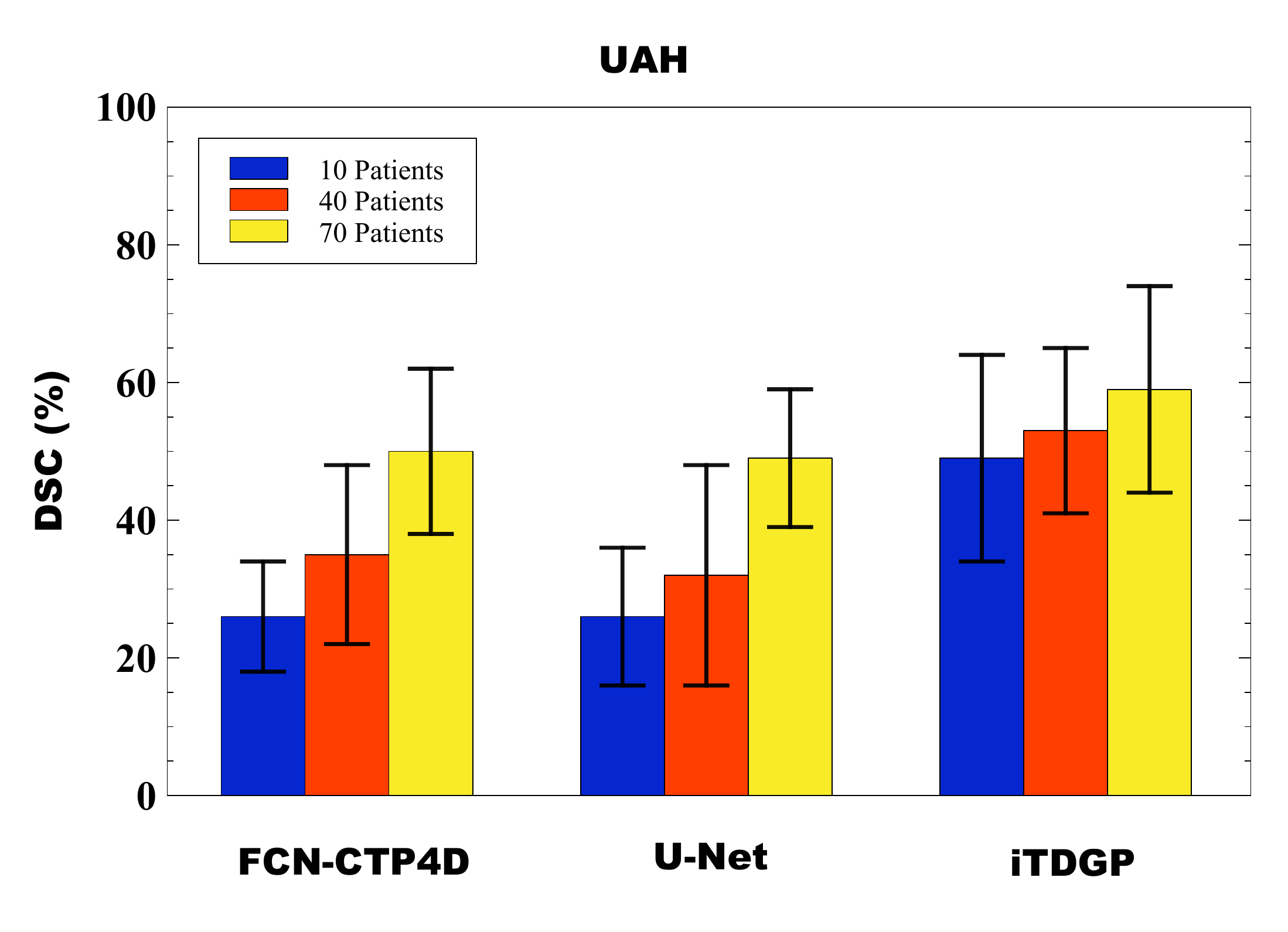}}
        \end{minipage}
		\caption{The DSC results that our proposed model achieves next to some selected state-of-the-art methods for learning from different sizes of training sets.}
		\label{fig:DCS}
	\end{minipage}
\end{center}
\vskip -0.15in        
\end{figure}

\subsection{TDGP vs. iTDGP}

In order to compare our TDGP and iTDGP, we conducted additional studies. Figure~\ref{fig:Visualize} presents the lesions predicted by TDGP and iTDGP, for two patients from each dataset. As mentioned earlier, iTDGP uses balanced batches for training and uses a postprocessing step, but TDGP does neither. 
Figure~\ref{fig:Visualize} suggests that iTDGP can identify AIS lesions more accurately and homogeneously by handling imbalanced classes and smoothing the predicted mask.


	

	


\begin{figure}[t]
	\begin{center}
		\includegraphics[width=0.7
		\textwidth]{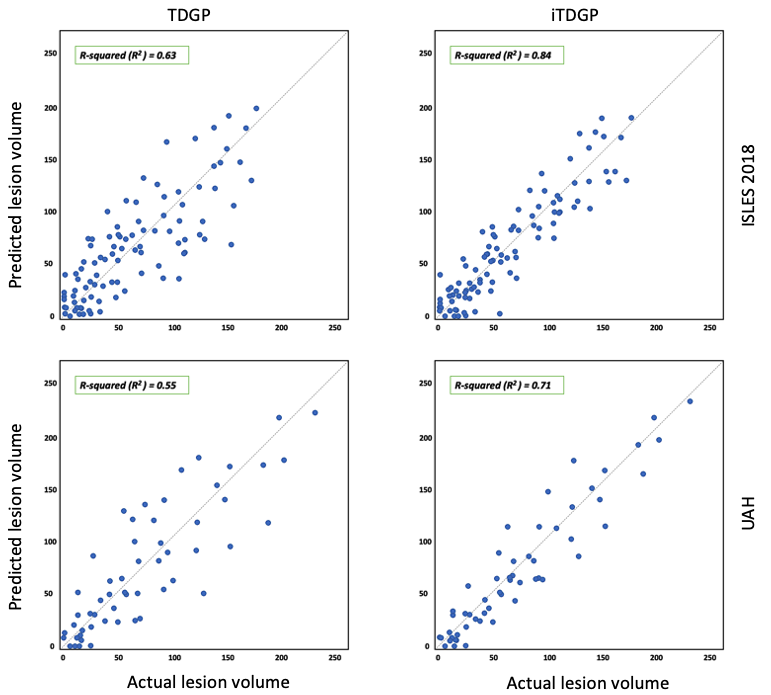}
	\end{center}
	\vskip -0.2in
	\caption{ Actual versus predicted lesion volumes(ml), by proposed TDGP --- that is iTDGP without
postprocessing and balanced batches --- (left column) and iTDGP (right column), for two datasets and using one-patient-out validation. }
	\label{fig:RScore}
\end{figure}

Another criterion that is commonly used to estimate the quality of an AIS lesion predictor is whether the predicted lesion is the same size (\ie, the same number of voxels) as the actual lesion. Figure~\ref{fig:RScore} shows a scatter plot of the true versus predicted volumes for all subjects of both ISLES and UAH datasets by using the proposed methods, where each dot on the plot corresponds to one patient in the dataset. The plots and associated R-squared (R2) values suggest that the proposed method can improve the accuracy of volume prediction of AIS lesions along with AIS lesion appearance prediction.   

\subsection{Small Dataset Effect}
In order to evaluate our proposed iTDGP in terms of its robustness in training on a small number of training examples, we simply compared it with two widely used deep neural networks,  FCN and U-Net. 
Figure~\ref{fig:DCS} compares the DSC score between our iTDGP, FCN-CTP4D, and U-Net on different dataset sizes. Since iTDGP uses voxel-level classification, it could technically augment the dataset so that the model could be learned with fewer patients.
Figure~\ref{fig:DCS} shows that our iTDGP model produces a better performance on datasets with low patient numbers in comparison with other methods. Moreover, when iTDGP is compared to the other methods, there is less difference between the DSCs obtained with the small and big datasets with iTDGP than with the others. This suggests that iTDGP can be more robust against the training set size problem than other methods.
The problem with small datasets is also that they contain a low number of voxels in the target class (which is relatively small). To learn, however, iTDGP uses balanced batches rather than the entire dataset at once. By using this technique, we can partially solve the problem of generalization over small sample size.

\section{Conclusion}\label{sec:conclusion}
This paper presents a new prediction model, called imbalanced Temporal Deep Gaussian Process (iTDGP), that can learn to predict AIS lesions based on the $4D$ CTP time series. This iTDGP extends the standard DGP approach to address two main drawbacks of the previous AIS lesion prediction models. First, it can effectively extract the temporal information of the CTP time series to predict the future AIS lesion. Secondly, iTDGP can address the imbalanced class problem (healthy versus lesion voxels) that is a critical barrier to improving AIS lesion prediction accuracy. Further, iTDGP uses a postprocessor to correct misclassified brain regions to improve lesion prediction accuracy. By comparing the obtained DSC by our method with previous works using both ISLES 2018 and UAH datasets, it was confirmed that iTDGP achieves performance that is superior to other state-of-the-art AIS lesion predictors. We anticipate that the iTDGP prediction model will have strong practical applications in different time series analyses — such as temporal modeling of localized brain activity.

\paragraph{Data availability:}The ISLES challenge 2018~\cite{cereda2016benchmarking} dataset is publicly available. 
The UAH dataset that was generated and analyzed during the current study is not publicly available, however, it can be made available from the corresponding author upon reasonable request.

\bibliography{sample}










\end{document}